\documentclass[pre,aps,preprint,showpacs,preprintnumbers,amsmath,amssymb,secnumroman,eqsecnum]{revtex4}

\usepackage{graphicx}
\usepackage{dcolumn}
\usepackage{bm}

\newcommand{\beq}{\begin{equation}}
\newcommand{\eeq}{\end{equation}}
\newcommand{\beqa}{\begin{eqnarray}}
\newcommand{\eeqa}{\end{eqnarray}}
\newcommand{\vc}[1]{\mbox{\boldmath $#1$}}

\newcommand{\vol}[1]{{\bf #1}}

\newcommand{\du}[1]{{\bf\sf #1}}

\begin{document}


\title{Collinear swimmer propelling a cargo sphere at low Reynolds number}

\author{B. U. Felderhof}

 \email{ufelder@physik.rwth-aachen.de}
\affiliation{Institut f\"ur Theorie der Statistischen Physik \\ RWTH Aachen University\\
Templergraben 55\\52056 Aachen\\ Germany\\
}%

\date{\today}

\begin{abstract}
The swimming velocity and rate of dissipation of a linear chain consisting of two or three little spheres and a big sphere is studied on the basis of low Reynolds number hydrodynamics. The big sphere is treated as a passive cargo, driven by the tail of little spheres via hydrodynamic and direct elastic interaction. The fundamental solution of Stokes' equations in the presence of a sphere with no-slip boundary condition, as derived by Oseen, is used to model the hydrodynamic interactions between the big sphere and the little spheres.
\end{abstract}

\pacs{47.15.G-, 47.63.mf, 47.63.Gd, 45.40.Ln}
\maketitle
\section{\label{I}Introduction}

The theory of swimming of microorganisms in a viscous fluid is based on low Reynolds number hydrodynamics \cite{1}. Inertia can be neglected and the relatively simple time-independent Stokes equations can be used. Nonetheless the mechanism of swimming is subtle and the study of the phenomenon has led to much fascinating theoretical work \cite{2}-\cite{4}.

Many microorganisms have the form of a large passive head propelled by a flagellum or tail. In the following we study the swimming of such structures for a simple model system, in which the tail is composed of little spheres which can move relative to each other and push or pull the head, taken to be a sphere of large radius. A model of this kind with two little spheres pushing a big one was first studied by Golestanian \cite{5}.

We consider in our explicit calculations two or three little spheres pushing a spherical head. The three-sphere swimmer by itself is effective \cite{6},\cite{7}, whereas a two-sphere swimmer, by itself in infinite fluid and with only collinear relative motions, does not swim at all according to Purcell's scallop theorem \cite{8}. In principle both longitudinal and transverse motions can be considered. In the longitudinal model the two or three little spheres and the big one are assumed to be aligned with motions and forces directed along the common axis. In the transverse model the spheres are again aligned but the forces are directed perpendicular to the axis. The swimming is a second order effect due to longitudinal or transverse waves propagating along the tail which couple to motion along the axis via the nonlinearity of hydrodynamic interactions. In numerical calculations the number of little spheres can be easily varied.

We employ a dynamic picture in which the forces acting on the individual spheres are decomposed into harmonic direct interactions and actuating forces summing to zero \cite{9}. The picture allows explicit study of the effect of the elastic direct interaction. A similar model of the flagellum has been studied in simulation by Lowe et al. \cite{10},\cite{11}. A flagellum with transverse motion is probably more relevant in nature. A complex active three-sphere longitudinal swimmer involving elastic forces has been studied by G\"unther and Kruse \cite{11A}. In these earlier calculations the hydrodynamic interactions were simplified as Oseen point interactions. Some experimental realizations of artificial swimming involve predominantly a flagellar motion \cite{11B},\cite{11C}. Other experiments on pumping and swimming have involved harmonic traps \cite{11D},\cite{11E}.

The present calculation is based on an approximation to the mobility matrix valid in the limit where the little spheres are at a mutual distance much larger than their radius, but much less than the radius of the big sphere. The hydrodynamic interaction between head and tail is calculated from the exact Green function of the Stokes equations for the geometry of a fixed sphere with no-slip boundary condition, derived by Oseen \cite{12}, as generalized to a moving sphere \cite{13}. We assume that the forces are sufficiently small that the bilinear theory of swimming can be employed. In Secs. II-IV we discuss the longitudinal model in detail. Corresponding results for the transverse model are more difficult to derive. In Sec. V we compare the results of the bilinear theory with a numerical solution of the equations of motion for the longitudinal model of three little spheres and a big one.

We find interesting behavior of swimming speed and rate of dissipation as functions of the ratio of the radius of the big sphere to that of the little spheres. The calculations show that estimates of the swimming speed based on Stokes' law \cite{14} and so-called resistive force theory \cite{15}-\cite{18} do not take proper account of hydrodynamic interactions and lead to an incorrect dependence of the swimming speed on the radius of the big sphere.

\section{\label{II}Collinear longitudinal swimmer}

In the following we consider a collinear swimmer consisting of $N$ spheres  with centers located sequentially on the $x$ axis of a Cartesian system of coordinates and linked together by nearest neighbor harmonic interactions. Specifically we consider a bead-spring chain for which the first $N-1$ beads have the same radius $a$, and the last one, labeled $N$, has radius $b$. The first $N-1$ spheres will jointly be called the tail or flagellum, and the last sphere will be called the head. We are particularly interested in the limiting behavior for $b>>a$. In Fig. 1 we show as an example a chain at rest consisting of three beads and a head of radius $b=5a$.

The chain is immersed in a viscous incompressible fluid of shear viscosity $\eta$ and of infinite extent. The fluid flow velocity $\vc{v}$ and the pressure $p$ are assumed to satisfy the steady-state Stokes equations,
\begin{equation}
\label{2.1}\eta\nabla^2\vc{v}-\nabla p=0,\qquad\nabla\cdot\vc{v}=0.
\end{equation}
A no-slip condition is assumed to hold on the surface of each sphere. The flow velocity and pressure are determined instantaneously by the sphere velocities.

The swimming behavior will be studied in the framework of the bilinear theory of swimming, in which the swimming velocity and the rate of dissipation are evaluated to second order in the amplitude of displacements of the spheres from their equilibrium positions. The displacements are assumed to vary harmonically in time with frequency $\omega$. The positions of centers correspond to an $N-1$-dimensional vector of relative positions $\du{r}=(r_1,...,r_{N-1})$ given by
\begin{equation}
\label{2.2}r_1=x_2-x_1,...,r_{N-1}=x_N-x_{N-1}.
\end{equation}
The equilibrium positions correspond to relative positions $\du{r}_0$. The difference $\du{r}(t)-\du{r}_0$ is assumed to vary harmonically in time $t$,
\begin{equation}
\label{2.3}\du{r}(t)-\du{r}_0=\hat{\vc{\xi}}(t)=\varepsilon[\hat{\vc{\xi}}_s\sin\omega t+\hat{\vc{\xi}}_c\cos\omega t],
\end{equation}
with dimensionless amplitude factor $\varepsilon$ and with constant vectors $\hat{\vc{\xi}}_s$ and $\hat{\vc{\xi}}_c$. The hat indicates that the vector has $N-1$ components, labeled $(1,...,N-1)$.

We restrict attention to the mean swimming velocity and the mean rate of dissipation, with the average performed over a period $T=2\pi/\omega$. We have shown elsewhere \cite{19} that these quantities can be expressed as
\begin{equation}
\label{2.4}\overline{U}_{sw}=\varepsilon^2\overline{U}_{sw2}+O(\varepsilon^3),\qquad\overline{\mathcal{D}}=\varepsilon^2\overline{\mathcal{D}}_2+O(\varepsilon^3),
\end{equation}
with second order terms
\begin{equation}
\label{2.5}\overline{U}_{sw2}=\frac{1}{2}\omega
a(\hat{\vc{\xi}}^c|\du{B}|\hat{\vc{\xi}}^c),\qquad\overline{\mathcal{D}}_2=\frac{1}{2}\eta\omega^2a^3(\hat{\vc{\xi}}^c|\du{A}|\hat{\vc{\xi}}^c),
 \end{equation}
with the complex dimensionless vector
 \begin{equation}
\label{2.6}\hat{\vc{\xi}}^c=\frac{1}{a}(\hat{\vc{\xi}}_c+i\hat{\vc{\xi}}_s),
 \end{equation}
the scalar product
  \begin{equation}
\label{2.7}(\hat{\vc{\xi}}^c|\hat{\vc{\eta}}^c)=\sum^{N-1}_{j=1}\hat{\xi}_j^{c*}\cdot\hat{\eta}^c_j,
 \end{equation}
and $(N-1)\times (N-1)$-dimensional hermitian matrices $\du{A}$ and $\du{B}$. We normalize the vector $\hat{\vc{\xi}}^c$ such that
  \begin{equation}
\label{2.8}(\hat{\vc{\xi}}^c|\hat{\vc{\xi}}^c)=1.
 \end{equation}
For present purposes it is convenient to use $a$, rather than $b$, as the fundamental length scale.

The matrices $\du{A}$ and $\du{B}$ depend on the geometric structure, as specified by $N$ sphere radii $\vc{a}=(a_1,...,a_N)$ and the relative equilibrium configuration $\du{r}_0=(r_{01},...,r_{0,N-1})$, and can be evaluated from the $N\times N$ translational mobility matrix $\vc{\mu}(\du{r})$. A more complete notation would be $\du{A}(\vc{a},\du{r}_0)$ and $\du{B}(\vc{a},\du{r}_0)$ with the dependence on parameters indicated explicitly. An approximation to the mobility matrix yields corresponding approximate matrices $\du{A}$ and $\du{B}$. In the following we use an approximation based on the assumption that the radius $a$ of the first $N-1$ spheres is much less than their mutual distances, the separation from sphere $N$, and the radius $b$ of sphere $N$.

A particular swimming stroke, as specified by a choice of the vector $\hat{\vc{\xi}}^c$, leads to a value of the mean swimming velocity $\overline{U}_{sw2}$ and the power $\overline{\mathcal{D}}_{2}$ as given by Eq. (2.5). By variation of the vector $\hat{\vc{\xi}}^c$ we can optimize the mean swimming velocity for fixed power. The optimization of the quadratic form $(\hat{\vc{\xi}}^c|\du{B}|\hat{\vc{\xi}}^c)$ at constant $(\hat{\vc{\xi}}^c|\du{A}|\hat{\vc{\xi}}^c)$ leads with a Lagrange multiplier $\lambda$ to a generalized eigenvalue problem
 \begin{equation}
\label{2.9}\du{B}\hat{\vc{\xi}}^c=\lambda\du{A}\hat{\vc{\xi}}^c.
 \end{equation}
 The maximum eigenvalue corresponds to optimal swimming. We define the dimensionless efficiency as \cite{20},\cite{21}
 \begin{equation}
\label{2.10}E_T=\eta\omega a^2\frac{|\overline{U}_{sw2}|}{\overline{\mathcal{D}}_2}.
\end{equation}
The maximum efficiency corresponds to the maximum eigenvalue as $E_{Tmax}=\lambda_{max}$.

In the following we compare the maximum efficiency with the efficiency for a limited class of strokes corresponding to the assumption that the $N$-th sphere is passive, as defined below. In general we assume that the
relative displacement vector $\hat{\vc{\xi}}^c$ is determined by actuation forces $\du{E}=(E_1,...,E_N)$ acting on each of the spheres with the constraint
 \begin{equation}
\label{2.11}\sum^N_{j=1}E_j=0,
\end{equation}
so that the total force vanishes. The force acting on sphere $j$ is the sum of the applied force $E_j$ and the direct interaction force. We assume harmonic interactions so that the first $N-1$ forces may be expressed as
 \begin{equation}
\label{2.12}\hat{\du{F}}=\hat{\du{E}}+\du{K}\cdot\hat{\vc{\xi}},
\end{equation}
where $\hat{\du{E}}=(E_1,...,E_{N-1})$ represents the first $N-1$ actuating forces and $\du{K}$ is a real matrix corresponding to the elastic interaction forces. The force $F_N$ acting on the head follows from the condition Eq. (2.11) and from Newton's third law for the elastic interaction. We note that the condition (2.11) implies that there are at least two actuating forces. This excludes the so-called one-armed swimmer \cite{22}. In our view such a body should not be called a swimmer, since the net force exerted on the fluid does not vanish at all times.

Passivity of the head is expressed by the condition $E_N=0$. It follows from Eq. (2.11) that this imposes a constraint on the first $N-1$ actuating forces. As a consequence, under this condition the maximum efficiency is less than the maximum eigenvalue $\lambda_{max}$.

In order to perform explicit calculations the mobility coefficients $\mu_{jk}(\du{r}_0)$ must be known. In our calculation we use an expression for the mobility matrix $\vc{\mu}$ which is exact in the limit of small radius $a$. The expression involves the Green tensor solution of the Stokes equations Eq. (2.1) in the presence of only the big sphere centered at $(x_N,0,0)$, as given by Oseen \cite{12}. The mobility matrix takes the form \cite{13}
\begin{equation}
\label{2.13}\vc{\mu}=\frac{1}{8\pi\eta}\left(\begin{array}{cc}M_{\alpha\beta}&M_{\alpha N}
\\M_{N\alpha}&M_{NN}
\end{array}\right),\qquad (\alpha,\beta)=(1,...,N-1),
\end{equation}
with a symmetric $(N-1)\times (N-1)$ matrix $\du{M}$ and with $M_{\alpha N}=M_{N\alpha}$, so that the $N\times N$ matrix $\vc{\mu}$ is symmetric, as it should by reciprocity \cite{1}. The explicit values are \cite{13}
\begin{eqnarray}
\label{2.14}M_{\alpha\beta}&=&M^s_{\alpha}\delta_{\alpha\beta}+M^d_{\alpha\beta}(1-\delta_{\alpha\beta}),\nonumber\\
M_{\alpha N}&=&
M_{N\alpha}=M^{St}(\rho_\alpha),\qquad M_{NN}=\frac{4}{3b},
\end{eqnarray}
with distance $\rho_\alpha=x_N-x_\alpha$ and coefficients given by
\begin{eqnarray}
\label{2.15}M^s_{\alpha}&=&\frac{4}{3a_\alpha}-\frac{b^3}{3\rho_\alpha^6}\frac{15\rho^4_\alpha-7b^2\rho^2_\alpha+b^4}{\rho^2_\alpha-b^2},\qquad
M^{St}(\rho_\alpha)=\frac{2}{\rho_\alpha}-\frac{2b^2}{3\rho_\alpha^3},\nonumber\\
M^d_{\alpha\beta}&=&\frac{2}{|x_\alpha-x_\beta|}+\frac{3}{4}bM^{St}(\rho_\alpha)M^{St}(\rho_\beta)
-b\frac{3\rho_\alpha^2\rho_\beta^2-b^2(\rho_\alpha^2+4\rho_\alpha \rho_\beta+\rho_\beta^2)+3b^4}{(\rho_\alpha\rho_\beta-b^2)^3},
\end{eqnarray}
where in the present case $a_\alpha=a$. Higdon has used Oseen's Green tensor in the framework of flagellar hydrodynamics \cite{23},\cite{24}.

The mechanism of swimming corresponding to the mobility matrix in Eq. (2.13) is the same as for a model with Oseen point interactions between the individual spheres. It is based on the nonlinear dependence of the mobility matrix on relative distances \cite{19}. The particular form (2.15) allows investigation of the limit where the radius of sphere $N$ is large compared to other length scales.

\section{\label{III}Actuating forces and direct interactions}

We consider first the general relation between actuating forces and relative displacements. This is determined by both hydrodynamic and elastic interactions. We consider actuating forces varying harmonically in time and put
 \begin{equation}
\label{3.1}E_\alpha(t)=\pi\eta a^2\omega(e_{\alpha s}\sin\omega t+e_{\alpha c}\cos\omega t),\qquad \alpha=1,...,N-1,
\end{equation}
with dimensionless factors $e_{\alpha s},e_{\alpha c}$. The force $E_N(t)$ is determined by Eq. (2.11). The complex actuating force amplitudes are
 \begin{equation}
\label{3.2}E_{\alpha\omega}^c=\pi\eta a^2\omega(e_{\alpha c}+ie_{\alpha s}),\qquad \alpha=1,...,N-1,
\end{equation}
with
 \begin{equation}
\label{3.3}E_\alpha(t)=\mathrm{Re}E_{\alpha\omega}^ce^{-i\omega t}.
\end{equation}
The corresponding displacement vector $\hat{\vc{\xi}}^c$ is found from the linearized equations of Stokesian dynamics.

The first order equations of motion in relative coordinate space are given by \cite{19}
  \begin{equation}
\label{3.4}\frac{d\du{r}}{dt}=\frac{d\hat{\vc{\xi}}}{dt}=\du{L}\cdot\hat{\du{F}},
 \end{equation}
 with relative mobility matrix
   \begin{equation}
\label{3.5}\du{L}=\du{S}\hat{\vc{\mu}}\big|_0,
 \end{equation}
 where the $(N-1)\times(N-1)$ matrices $\du{S}$ and $\hat{\vc{\mu}}$ can be evaluated from the mobility matrix $\vc{\mu}(\du{r})$ and the subscript 0 indicates that these are evaluated at $\du{r}_0$.
 By use of Eq. (2.12) we find that for harmonically oscillating forces the complex amplitude vector is given by
    \begin{equation}
\label{3.6}\hat{\vc{\xi}}^c_\omega=a^{-1}\big[-i\omega\du{I}-\du{L}\du{K}\big]^{-1}\du{L}\cdot\hat{\du{E}}^c_\omega.
 \end{equation}
 This can be used to calculate the mean swimming velocity and the mean rate of dissipation from Eqs. (2.5) in terms of the actuating forces. The normalization Eq. (2.8) determines a corresponding normalization of the actuating forces $\hat{\du{E}}^c_\omega$. From Eq. (3.2) we define
     \begin{equation}
\label{3.7}\hat{\vc{E}}^c_\omega=\pi\eta a^2\omega\hat{\vc{e}}^c,
 \end{equation}
 and write the relation Eq. (3.6) in the form
    \begin{equation}
\label{3.8}\hat{\vc{\xi}}^c=\du{X}\cdot\hat{\vc{e}}^c,
 \end{equation}
 with a dimensionless matrix $\du{X}$.

 If in addition we impose the condition that the big sphere is a passive cargo by requiring $E_N=0$, then there are only $N-2$ independent actuating forces, for which we can take, for example, $E_1,...,E_{N-2}$. The big sphere does exert a hydrodynamic force on the fluid, but only due to the elastic interaction with its neighboring sphere labeled $N-1$. We call the harmonic link between spheres $N-1$ and $N$ the attachment. The force on the big sphere is transmitted via the attachment.

 Specifically we assume that in the equilibrium situation the little spheres of the tail are equidistant with distance $d$ between centers and that the center of the big sphere is at distance $b+d$ from the $N-1$-th sphere. Moreover we assume that the strength of the harmonic interaction is characterized by a single elastic constant $k$. Then for $N=4$ the matrix $\du{K}$ takes the form
\begin{equation}
\label{3.9}\du{K}=k\left(\begin{array}{ccc}1&0&0\\
-1&1&0\\0&-1&1
\end{array}\right),\qquad (N=4).
\end{equation}
The corresponding form for $N=3$ and $N>4$ is obvious.

The matrices $\du{A}$ and $\du{B}$ are determined by the hydrodynamic interactions only. The explicit expressions calculated by use of Eq. (2.14) are quite complicated, but numerical values are easily obtained. For $N=3$ the equilibrium configuration takes the form $\du{r}_0=(d,b+d)$, and for $N=4$ the configuration is $\du{r}_0=(d,d,b+d)$. We present analytic results in the limit $a<<d<<b$. For $N=3$ the matrices $\du{A}$ and $\du{B}$ take the asymptotic form
\begin{equation}
\label{3.10}\du{A}_{as}=6\pi\left(\begin{array}{cc}1&1\\
1&2
\end{array}\right),\qquad\du{B}_{as}=\frac{15a^3}{4b^3}\left(\begin{array}{cc}0&-i\\
i&0
\end{array}\right),\qquad (N=3).
\end{equation}
In the same limit for $N=4$ the matrix $\du{A}$ takes the asymptotic form
\begin{equation}
\label{3.11}\du{A}_{as}=6\pi\left(\begin{array}{ccc}1&1&1\\
1&2&2\\1&2&3
\end{array}\right),\qquad (N=4),
\end{equation}
and the matrix $\du{B}$ becomes
 \begin{equation}
\label{3.12}\du{B}_{as}=\frac{3ia^3}{40b^3}\left(\begin{array}{ccc}0&-168&-183\\
168&0&-65\\183&65&0
\end{array}\right),\qquad (N=4).
\end{equation}

We characterize the stiffness of the swimmer by the dimensionless number $\sigma$ defined by
\begin{equation}
\label{3.13}\sigma=\frac{k}{\pi\eta a\omega}.
\end{equation}
It turns out that for a given set of actuating forces $(E_1,...,E_{N-1})$ both the swimming speed and the rate of dissipation depend on this dimensionless parameter. For fixed parameter $\sigma$ we can optimize the efficiency with respect to the actuating forces. Subsequently we can look for the value of $\sigma$ which yields the largest efficiency.

\section{\label{IV}Three- and four-sphere chains}

As examples we consider in the following chains consisting of three or four spheres, denoted as 3-chains and 4-chains. If the head is passive, as expressed by the cargo constraint $E_N=0$, we denote the chains as $3C$-chain and $4C$-chain, respectively.

We express the mean swimming speed and the mean rate of dissipation as
\begin{equation}
\label{4.1}|\overline{U}_{sw2}|=\omega a\hat{U}_2,\qquad\overline{\mathcal{D}}_2=\eta \omega^2a^3\hat{D}_2
\end{equation}
with dimensionless functions $\hat{U}_2$ and $\hat{\mathcal{D}}_2$ which depend on the ratios $d/a,b/a$ and on the stiffness $\sigma$. From Eq. (2.5) we find
\begin{equation}
\label{4.2}\hat{U}_2=\frac{1}{2}|(\hat{\vc{\xi}}^c|\du{B}|\hat{\vc{\xi}}^c)|,
\qquad\hat{\mathcal{D}}_2=\frac{1}{2}(\hat{\vc{\xi}}^c|\du{A}|\hat{\vc{\xi}}^c).
 \end{equation}
We call $\hat{U}_2$ with the normalization Eq. (2.8) the reduced speed and $\hat{\mathcal{D}}_2$ the reduced power.
We shall consider in particular the asymptotic limit $a<<d<<b$ and the case $d=5a$, $b=10a$.

For a 3-chain there are two independent actuating forces $E^c_1$ and $E^c_2$. The third force is given by $E^c_3=-E^c_1-E^c_2$. The optimum value of the reduced speed for given reduced power is found from the solution of the eigenvalue problem Eq. (2.9). The eigenvector with maximum eigenvalue in the limit $a<<d<<b$ is found from Eq. (3.10) as
\begin{equation}
\label{4.3}\hat{\vc{\xi}}^c_0=\frac{1}{\sqrt{6}}(2,-1+i),\qquad\lambda_{max}=\frac{5a^3}{8\pi b^3},\qquad (a<<d<<b).
\end{equation}
The subsript 0 is used here and in the following to indicate the eigenvector with maximum eigenvalue.
The corresponding reduced speed and power are
\begin{equation}
\label{4.4}\hat{U}_{2as}=\frac{5a^3}{4b^3},
\qquad\hat{\mathcal{D}}_{2as}=2\pi,
 \end{equation}
 and the efficiency is $E_{Tas}=5a^3/(8\pi b^3)=0.199a^3/b^3$. The actuating forces $(E^c_1,E^c_2)$ necessary to achieve the optimum can be evaluated from $\hat{\vc{\xi}}^c_0$ by inversion of Eq. (3.8). These forces depend on the elasticity of the chain.

For a $3C$-chain the actuating forces are related by $E^c_2=-E^c_1$, since $E^c_3=0$ and Eq. (2.11) must be satisfied. According to a general argument the $3C$-chain with $\sigma=0$ cannot swim \cite{13}. We can see this in the bilinear theory from the fact that the matrix $\du{B}$ takes the form
  \begin{equation}
\label{4.5}
\du{B}=\left(\begin{array}{cc}0&-i\beta_{12}\\
i\beta_{12}&0
\end{array}\right),
 \end{equation}
 with real $\beta_{12}$. For $\sigma=0$ the amplitude vector $\hat{\vc{\xi}}^c$ is proportional to $(1,-1)$, and the matrix element $(1,-1).\du{B}.(1,-1)$ vanishes. Golestanian's calculation \cite{5} was correct, but he did not notice that the constraint $F_3=0$ entails the phase relation $F_2=-F_1$, causing the representative point to move back and forth along a curve in $r_1r_2$ space, leading to vanishing swimming velocity.

 The $3C$-chain can swim only due to the presence of direct interaction forces, with the vector $\vc{\xi}^c$ determined by Eq. (3.8). For $N=3$ and $\sigma>0$ there is no possibility of optimization and Eq. (3.8) yields the vector $\hat{\vc{\xi}}^c$ for $\hat{\vc{e}}^c=(1,-1)$. In the limit $a<<d<<b$ we find for the corresponding reduced speed and power
  \begin{equation}
\label{4.6}\hat{U}^C_{2as}=\frac{45\sigma a^3}{2(180+\sigma^2)b^3},
\qquad\hat{\mathcal{D}}^C_{2as}=3\pi\frac{72+\sigma^2}{180+\sigma^2}.
 \end{equation}
The corresponding efficiency $E^C_{Tas}=\hat{U}^C_{2as}/\hat{\mathcal{D}}^C_{2as}$ is maximum at $\sigma_0=6\sqrt{2}\approx 8.485$. In Fig. 2 we show the behavior of the relative efficiency $(b^3/a^3)E^C_{Tas}$ as a function of $\sigma$. At the maximum
 \begin{equation}
\label{4.7}\hat{U}^C_{2as0}=\frac{3\sqrt{2}}{7}\hat{U}_{2as},
\qquad\hat{\mathcal{D}}^C_{2as0}=\frac{6}{7}\hat{\mathcal{D}}_{2as},
 \end{equation}
with efficiency  $E^C_{Tas0}=E_{Tas}/\sqrt{2}\approx 0.707E_{Tas}$.
In Fig. 2 we also show the behavior of the relative efficiency $(b^3/a^3)E^C_{T}$ for the case $d=5a,\;b=10a$. This shows a maximum at $\sigma_0=9.054$ with efficiency $E^C_{T0}=0.159E_{Tas}$. The maximum efficiency in this case is $E_T=0.261E_{Tas}=5\times 10^{-5}$, as found from the eigenvalue problem Eq. (2.9), so that $E^C_{T0}=0.608E_{T}$. In Fig. 3 we show the behavior of $(b^3/a^3)E^C_T$ for $d=5a,\;\sigma=10$ as a function of $b/a$.

For a 4-chain there are three independent actuating forces $E^c_1,E^c_2,$ and $E^c_3$.  The eigenvector with maximum eigenvalue of the eigenvalue problem Eq. (2.9) in the limit $a<<d<<b$ is given by
\begin{equation}
\label{4.8}\hat{\vc{\xi}}^c_0=(0.806,-0.266+0.499i,-0.153-0.086i),\qquad (a<<d<<b).
\end{equation}
with eigenvalue
\begin{equation}
\label{4.9}\lambda_{max}=0.700\frac{a^3}{b^3},\qquad (a<<d<<b).
\end{equation}
The corresponding reduced speed and power are
\begin{equation}
\label{4.10}\hat{U}_{2as}=4.598\frac{a^3}{b^3},
\qquad\hat{\mathcal{D}}_{2as}=6.569,
 \end{equation}
 and the efficiency is $E_{Tas}=0.700a^3/b^3$. For the case $d=5a,\;b=10a$ the eigenvector with maximum eigenvalue is given by
\begin{equation}
\label{4.11}\hat{\vc{\xi}}^c_0=(0.720,-0.121+0.598i,-0.311-0.112i),\qquad (d=5a,b=10a).
\end{equation}
with eigenvalue $\lambda_{max}=930\times 10^{-7}$.
The corresponding reduced speed and power are
\begin{equation}
\label{4.12}\hat{U}_{2}=0.000792,
\qquad\hat{\mathcal{D}}_{2}=8.523,
 \end{equation}
 and the efficiency equals $\lambda_{max}$.

In Fig. 4 we show the behavior of the relative maximum efficiency $(b^3/a^3)E^C_{Tas}$ of a $4C$-chain in the limit $a<<d<<b$ as a function of $\sigma$. The efficiency $E^C_{Tas}=\hat{U}^C_{2as}/\hat{\mathcal{D}}^C_{2as}$ is maximum at $\sigma_0=9.124$. At the maximum
the efficiency is $E^C_{Tas0}=0.698E_{Tas}$.
In Fig. 4 we also show the behavior of the relative maximum efficiency $(b^3/a^3)E^C_{Tmax}$ for the case $d=5a,\;b=10a$. This shows a maximum at $\sigma_0=6.271$ with efficiency $E^C_{T0}=0.132E_{Tas}$. In Fig. 5 we show the behavior of $(b^3/a^3)E^C_{Tmax}$ for $d=5a,\;\sigma=10$ as a function of $b/a$.

We compare the behavior found above with two conventional estimates of the swimming speed for given power consumption. Both lead to a dependence on the size of the cargo different from what we have found above. We consider a fixed value of the power consumption of the tail independent of the size $b$ of the cargo. In the first estimate \cite{14} Stokes' law is used to equate the power to $6\pi\eta b U^2$. This implies that the swimming speed $U$ decreases as $1/\sqrt{b}$ with the size of the cargo $b$. In the second estimate \cite{2}, as used in so-called resistive force theory \cite{15}-\cite{18}, it is assumed that the tail produces a propulsive force or thrust which is equated to the Stokes friction coefficient of the whole structure times the swimming speed. In our case the friction coefficient of the structure would be estimated as $6(N-1)\pi\eta a+6\pi\eta b$. Correspondingly the estimate suggests that the swimming speed decreases inversely with the size of the cargo. Both estimates are in conflict with our exact calculation which yields a speed proportional to $a^3/b^3$ in the limiting case $a<<d<<b$. We conclude that Stokes' law cannot be used for an estimate of the swimming speed. A complete calculation of swimming speed and rate of dissipation taking full account of hydrodynamic interactions, as performed here, is required.

\section{\label{V}Numerical solution}

It is of interest to compare the above analytical results with a numerical solution of the equations of Stokesian dynamics. In abbreviated form these read
\begin{equation}
\label{5.1}\frac{d\du{R}}{dt}=\vc{\mu}\cdot\du{F},
 \end{equation}
where in the present case the $N$-dimensional vector $\du{R}$ comprises the $N$ positions $(x_1,...,x_N)$ of the sphere centers on the $x$ axis, the $N\times N$ mobility matrix $\vc{\mu}$ depends on $\du{R}$, and the vector $\du{F}$ comprises the $N$ forces in $x$ direction. We have found in earlier work on the swimming of three equal-sized spheres that the bilinear theory provides a good approximation to swimming speed and power even for large amplitude strokes \cite{19}. Therefore the analytic results of the bilinear theory provide a useful guide to the actual behavior. We find that the same is true in the cargo problem studied here.

The forces are expressed as in Eq. (2.12) as
\begin{equation}
\label{5.2}\du{F}=\du{E}+\du{H}\cdot\vc{\xi},
 \end{equation}
 with actuating forces $\du{E}=(E_1,...,E_N)$, harmonic interaction matrix $\du{H}$ and displacements from equilibrium positions $\vc{\xi}=(\xi_1,...,\xi_N)$.
In particular we show numerical results for the $4C$-chain. Then the harmonic interaction matrix corresponding to Eq. (3.9) is given by
 \begin{equation}
\label{5.3}\du{H}=k\left(\begin{array}{cccc}-1&1&0&0\\
1&-2&1&0\\0&1&-2&1\\0&0&1&-1
\end{array}\right),\qquad (N=4).
\end{equation}
In the cargo problem $E_4=0$ and $E_3=-E_1-E_2$. We normalize such that $E_1=\varepsilon\eta\omega a^2$ at $t=0$, so that the applied force vector $\du{E}(t)$ takes the form
 \begin{equation}
\label{5.4}\du{E}(t)=\varepsilon\eta\omega a^2\mathrm{Re}(1,f+ig,-1-f-ig,0)e^{-i\omega t},
\end{equation}
with real $f$ and $g$ determined such that the efficiency $E_T$ of the bilinear theory is maximized. We consider a $4C$-chain with $d=5a,\;b=10a,\;\sigma=6.271$. In Fig. 6 we show the motion in the $r_1r_2$ plane for the first ten periods for $\varepsilon=100$, where $r_1=x_2-x_1,\;r_2=x_3-x_2$. The initial conditions have been chosen in accordance with Eq. (4.11). It is seen that the motion quickly tends to a limit cycle. It is evident that the stroke has large amplitude. In Fig. 7 we show the positions of the three little spheres during the last period.

The time-dependent swimming velocity $U_{sw}(t)$ and rate of dissipation $\mathcal{D}(t)$ are
 \begin{equation}
\label{5.5}U_{sw}(t)=\frac{1}{4}(U_1+U_2+U_3+U_4),\qquad\mathcal{D}(t)=F_1U_1+F_2U_2+F_3U_3+F_4U_4.
\end{equation}
The mean swimming velocity and power are calculated as time-averages over the last period. These quantities vary nearly quadratically as a function of the amplitude factor $\varepsilon$ in the range $0<\varepsilon<100$, close to the quadratic behavior of the bilinear theory. In Fig. 8 we show the mean rate of dissipation $\overline{\mathcal{D}}$ as a function of $\varepsilon$. The dependence is nearly exactly quadratic. In Fig. 9 we show the efficiency $E_T=\eta\omega a^2\overline{U}/\overline{\mathcal{D}}$ as a function of $\varepsilon$. The efficiency increases with increasing amplitude.

\section{\label{VIII}Discussion}

The explicit decomposition of the forces acting on the individual spheres into actuating forces and harmonic interaction forces provides a dynamic picture of the swimming motion \cite{9}. The separate calculation of swimming velocity and rate of dissipation allows optimization of the efficiency.

In the cargo problem it is tempting to use the idea of thrust and drag \cite{2},\cite{25}-\cite{29}, as in resistive force theory \cite{15}-\cite{18}. The thrust-drag paradigm is commonly used also in the theory of flying \cite{25},\cite{26},\cite{30}. As we have shown in Sec. IV, the corresponding estimate of the swimming velocity is in conflict with the behavior found in our model calculation. Our formulation in terms of a mobility matrix guarantees that proper and consistent account is taken of hydrodynamic interactions, at least in the framework of low Reynolds number hydrodynamics. The use of elastic direct interactions guarantees that proper account is taken of Newton's third law.

The dependence of the swimming efficiency on the ratio of radii $b/a$ shown in Figs. 2 and 4 appears simple, but the analytic dependence is actually quite complicated. The limiting behavior of the matrices $\du{A}$ and $\du{B}$, as shown, for example, in Eq. (3.9) for a longitudinal three-sphere swimmer, is simple, but defies a simple explanation. As shown elsewhere \cite{13}, in the case of collinear sedimentation of a sphere and two small particles the two-body Rotne-Prager approximation \cite{31} to the hydrodynamic interaction leads to the same asymptotic behavior of the sedimentation velocity for large ratio $b/a$ as for the mobility matrix we have used in Eqs. (2.15) and (5.1). We do not find this for swimming. In the equivalent of Eq. (3.9) calculated with the Rotne-Prager approximation we find the same matrix $\du{A}_{as}$, but a different pre-factor in the matrix $\du{B}_{as}$.

We have used an approximation to the mobility matrix which is valid if the radii of the little spheres are much smaller than mutual distances and than the radius of the big sphere. The approximation corresponds to the first few terms in a systematic expansion of the exact mobility matrix in powers of the corresponding ratios. The approximation allows explicit calculation, but is not necessary in principle. More complicated assemblies of rigid spheres with small mutual distances can be considered. The mobility matrix of such an assembly can in principle be evaluated by the method of multipole expansion \cite{32}-\cite{34}, though at the expense of more elaborate calculation. \\

We have restricted the analysis to the case of longitudinal motion. For transversely actuating forces qualitatively similar results can be derived. The analysis is technically more complicated, since motions both parallel and perpendicular to the axis must be considered, requiring a doubling of dimensions compared to the longitudinal case. For short chains of three or four spheres analytic results analogous to those for longitudinal swimming can be derived. More details, together with numerical results for longer chains, will be presented elsewhere.

$\vc{\mathrm{Acknowledgment}}$ I thank Dr. M. L. Ekiel-Je\.zewska for a critical reading of the manuscript and for helpful suggestions.

\newpage

\section*{Figure captions}

\subsection*{Fig. 1}
Schematic shape of longitudinal swimmer consisting of three beads and a cargo sphere.

\subsection*{Fig. 2}
Plot of the reduced efficiency $(b^3/a^3)E^C_T$ of the $3C$-chain with longitudinal linear motion as a function of the stiffness parameter $\sigma$ in the asymptotic limit $a<<d<<b$ (solid curve) and for $d=5a,\;b=10a$ (dashed curve).

\subsection*{Fig. 3}
Plot of the reduced efficiency $(b^3/a^3)E^C_T$ of the $3C$-chain with longitudinal linear motion as a function of the ratio $b/a$ for $d=5a$ and $\sigma=10$.

\subsection*{Fig. 4}
Plot of the reduced efficiency $(b^3/a^3)E^C_{Tmax}$ of the $4C$-chain with longitudinal linear motion as a function of the stiffness parameter $\sigma$ in the asymptotic limit $a<<d<<b$ (solid curve) and for $d=5a,\;b=10a$ (dashed curve).

\subsection*{Fig. 5}
Plot of the reduced efficiency $(b^3/a^3)E^C_{Tmax}$ of the $4C$-chain with longitudinal linear motion as a function of the ratio $b/a$ for $d=5a$ and $\sigma=10$.

\subsection*{Fig. 6}
Plot of the motion of a $4C$-chain with $d=5a,\;b=10a,\;\sigma=6.271$ in the $r_1r_2$ plane, where $r_1=x_2-x_1,\;r_2=x_3-x_2$, for ten periods with actuating forces as given by Eq. (5.3) with $\varepsilon=100$.

\subsection*{Fig. 7}
Plot of the positions $x_1(t),x_2(t),x_3(t)$ for the $4C$-chain during the last period of Fig. 5.

\subsection*{Fig. 8}
Plot of the mean rate of dissipation of a $4C$-chain with $d=5a,\;b=10a,\;\sigma=6.271$ as a function of the amplitude factor $\varepsilon$.

\subsection*{Fig. 9}
Plot of the efficiency $E_T$ of a $4C$-chain with $d=5a,\;b=10a,\;\sigma=6.271$ as a function of the amplitude factor $\varepsilon$.

\newpage

\newpage
\setlength{\unitlength}{1cm}
\begin{figure}
 \includegraphics{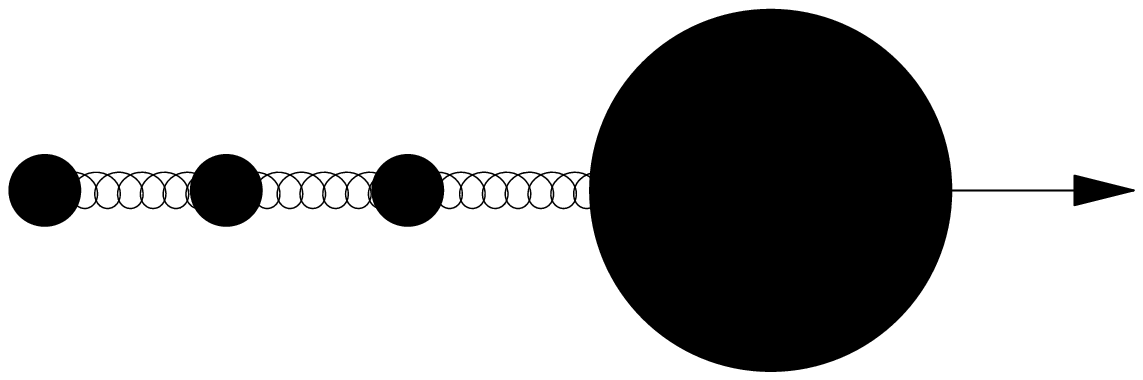}
   \put(-9.1,3.1){}
\put(-1.2,-.2){}
  \caption{}
\end{figure}
\newpage
\clearpage
\newpage
\setlength{\unitlength}{1cm}
\begin{figure}
 \includegraphics{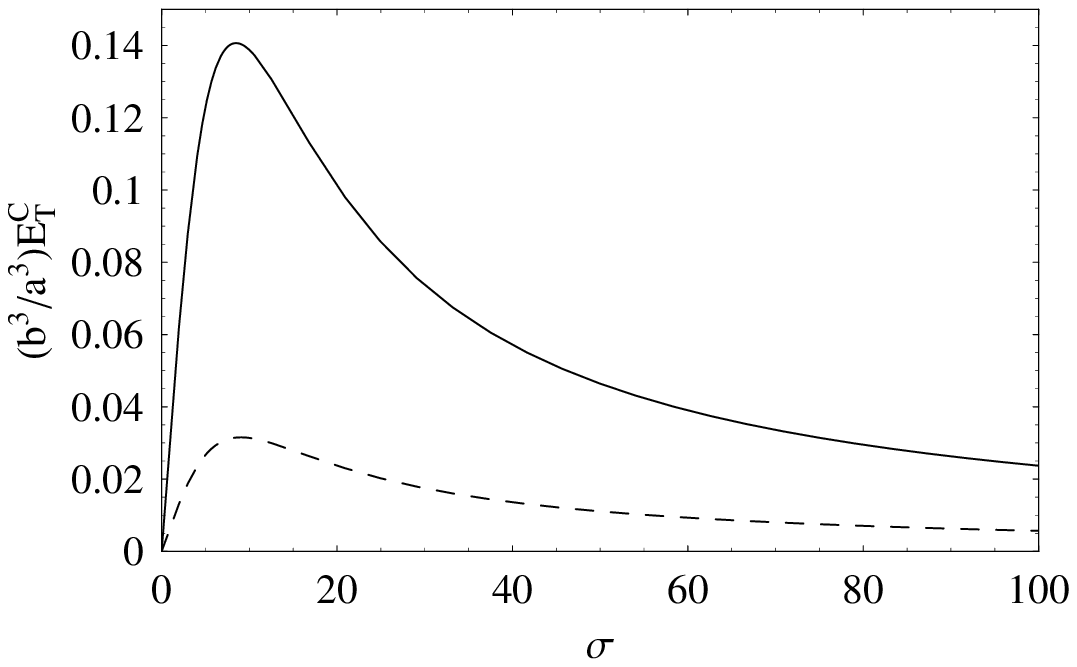}
   \put(-9.1,3.1){}
\put(-1.2,-.2){}
  \caption{}
\end{figure}
\newpage
\clearpage
\newpage
\setlength{\unitlength}{1cm}
\begin{figure}
 \includegraphics{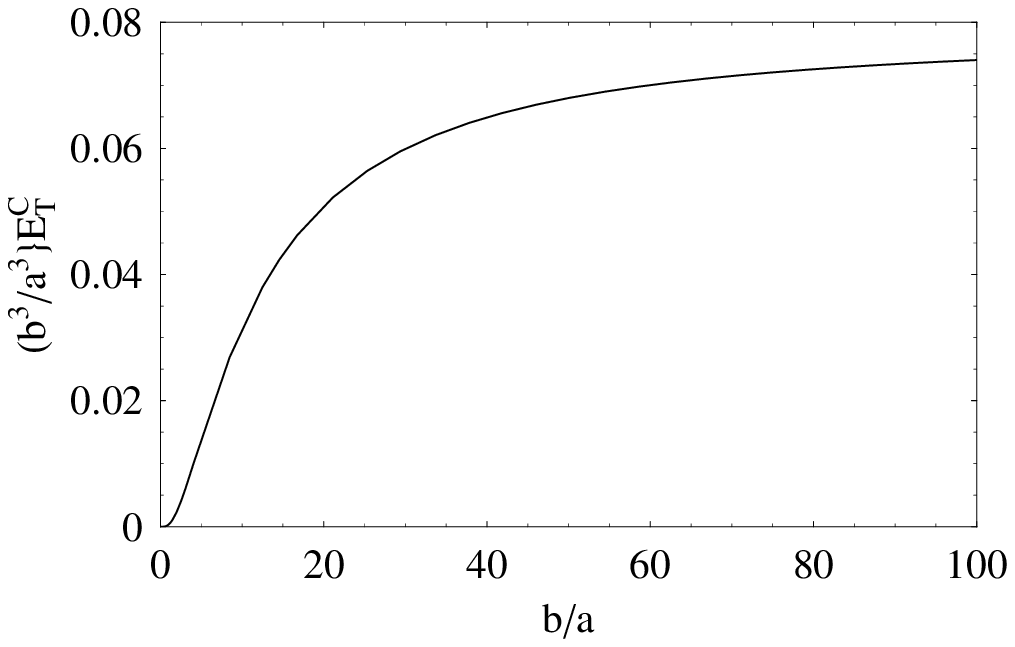}
   \put(-9.1,3.1){}
\put(-1.2,-.2){}
  \caption{}
\end{figure}
\newpage
\clearpage
\newpage
\setlength{\unitlength}{1cm}
\begin{figure}
 \includegraphics{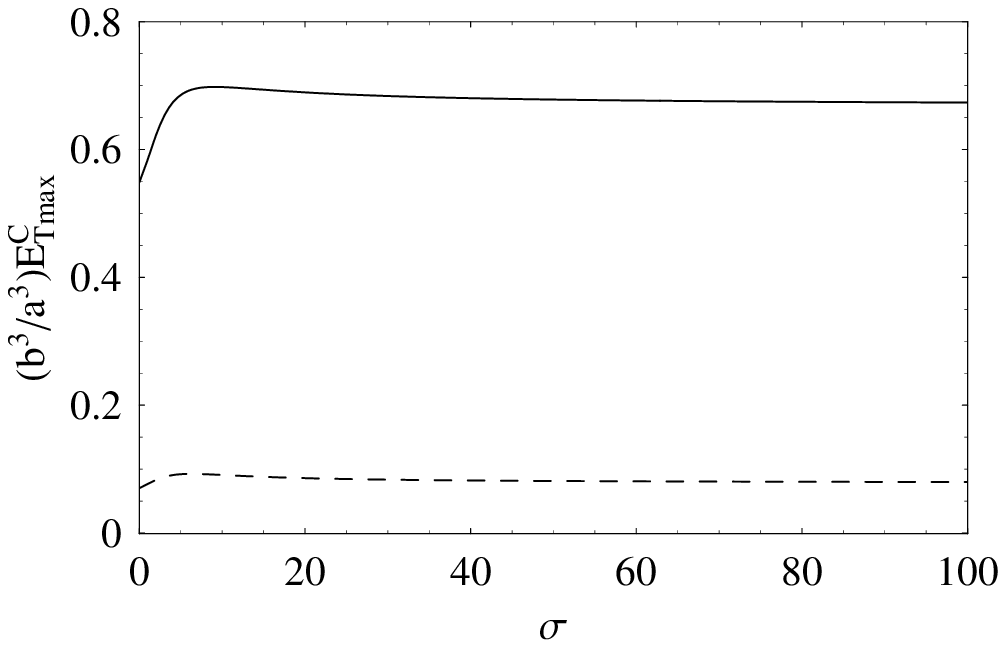}
   \put(-9.1,3.1){}
\put(-1.2,-.2){}
  \caption{}
\end{figure}
\newpage
\clearpage
\newpage
\setlength{\unitlength}{1cm}
\begin{figure}
 \includegraphics{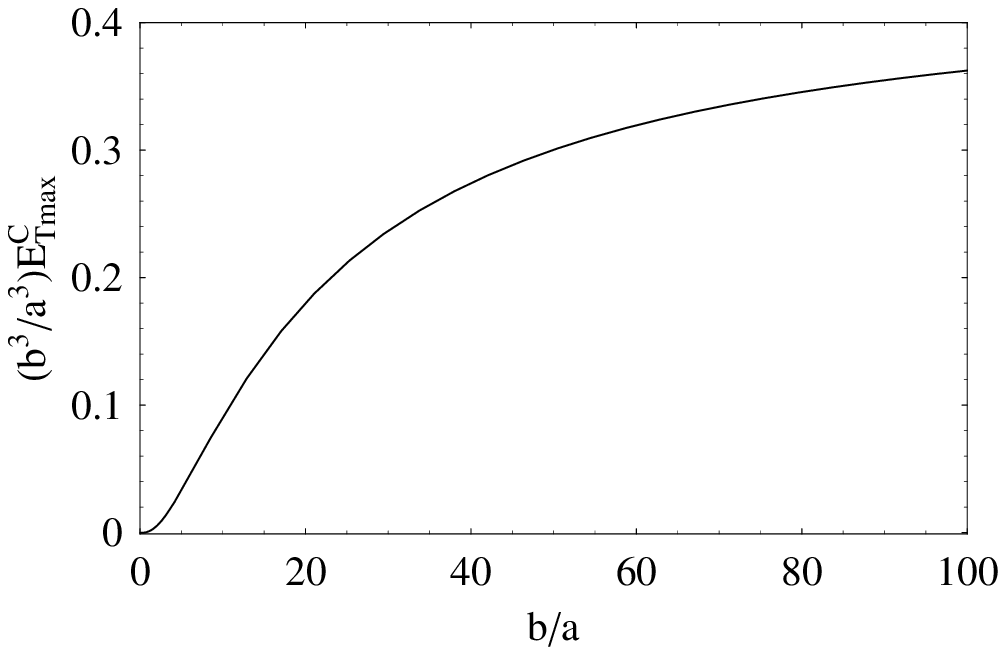}
   \put(-9.1,3.1){}
\put(-1.2,-.2){}
  \caption{}
\end{figure}
\newpage
\clearpage
\newpage
\setlength{\unitlength}{1cm}
\begin{figure}
 \includegraphics{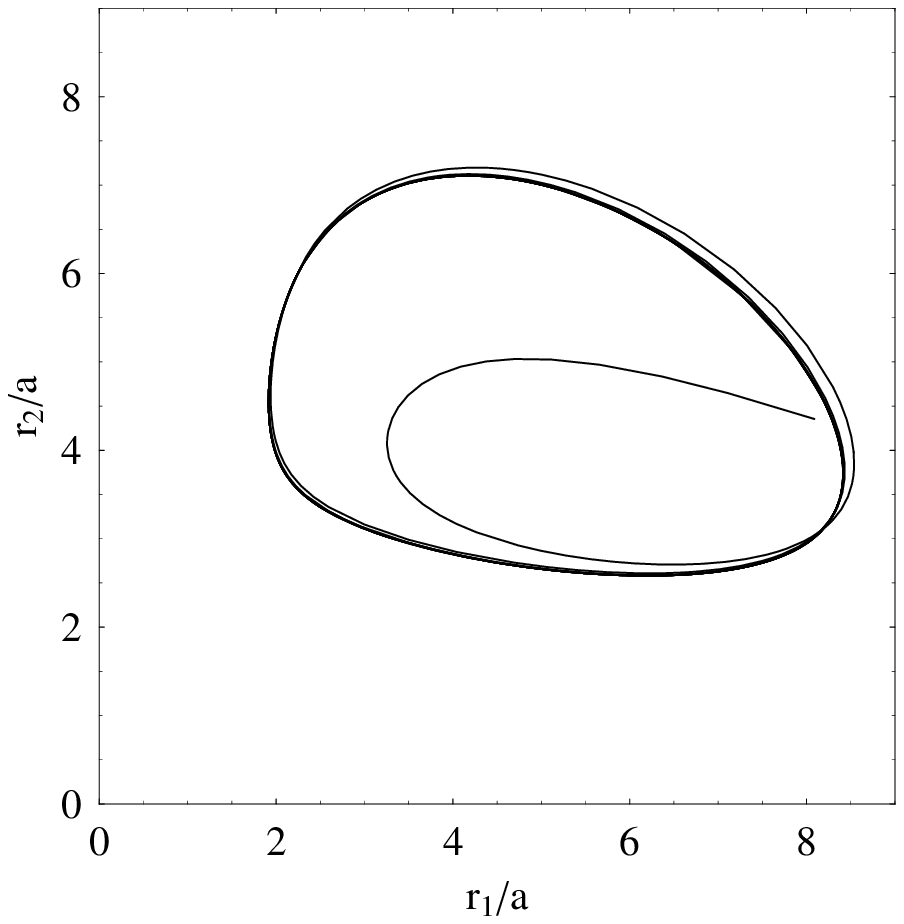}
   \put(-9.1,3.1){}
\put(-1.2,-.2){}
  \caption{}
\end{figure}
\newpage
\clearpage
\newpage
\setlength{\unitlength}{1cm}
\begin{figure}
 \includegraphics{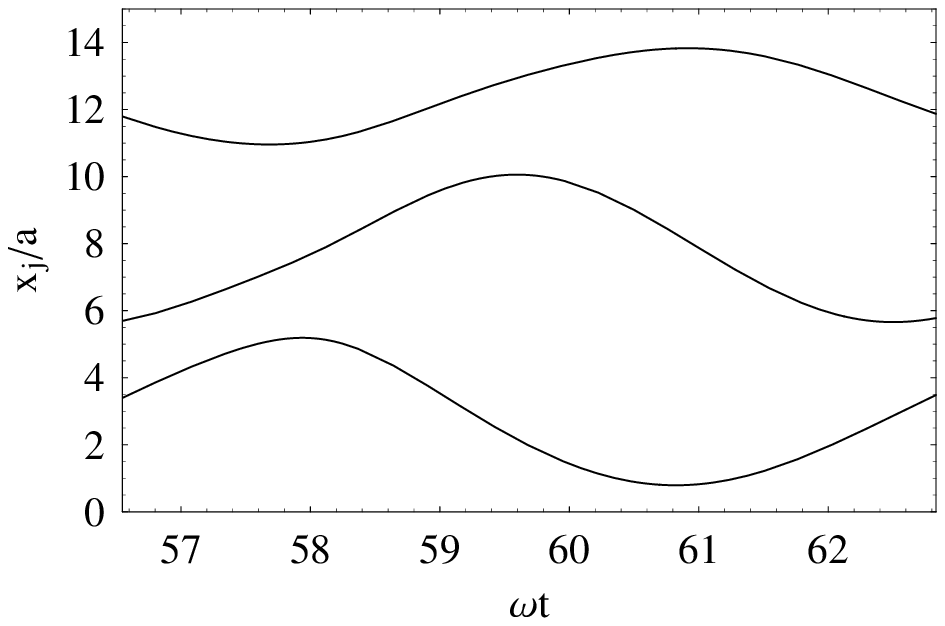}
   \put(-9.1,3.1){}
\put(-1.2,-.2){}
  \caption{}
\end{figure}
\newpage
\clearpage
\newpage
\setlength{\unitlength}{1cm}
\begin{figure}
 \includegraphics{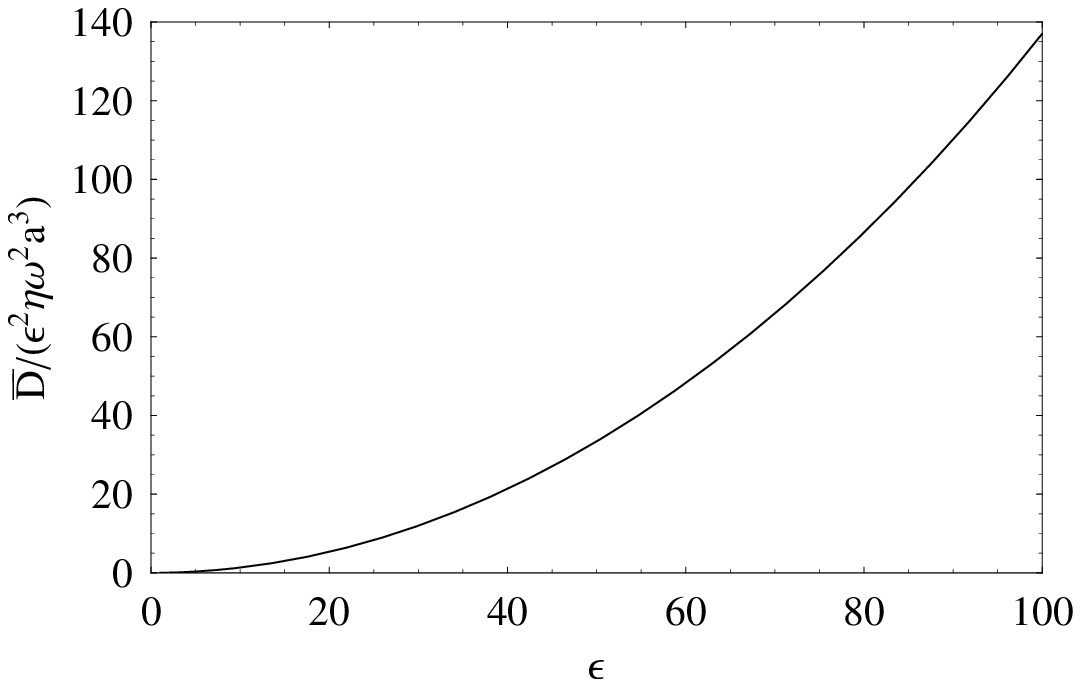}
   \put(-9.1,3.1){}
\put(-1.2,-.2){}
  \caption{}
\end{figure}
\newpage
\clearpage
\newpage
\setlength{\unitlength}{1cm}
\begin{figure}
 \includegraphics{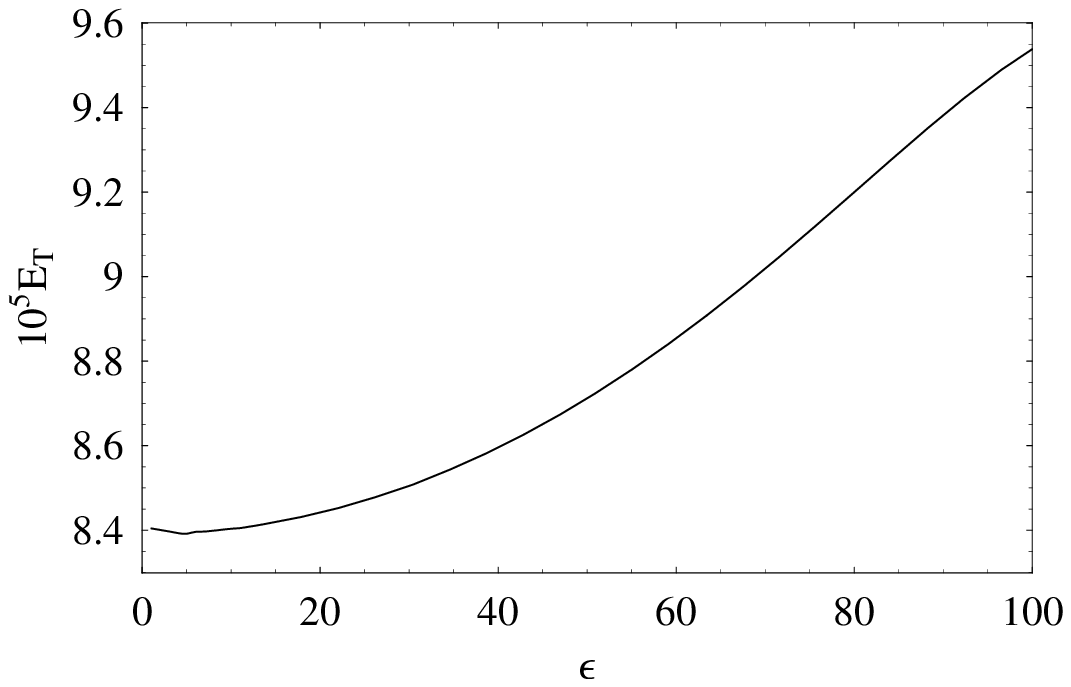}
   \put(-9.1,3.1){}
\put(-1.2,-.2){}
  \caption{}
\end{figure}
\newpage
\end{document}